\documentstyle[12pt]{article}

\def\ergs{{{\rm erg~s}^{-1}}}
\def\la{\mathrel{\mathpalette\fun <}}
\def\ga{\mathrel{\mathpalette\fun >}}
\def\fun#1#2{\lower3.6pt\vbox{\baselineskip0pt\lineskip.9pt
        \ialign{$\mathsurround=0pt#1\hfill##\hfil$\crcr#2\crcr\sim\crcr}}}

\begin{document}
\bibliographystyle{unsrt}

\begin{flushright} \small{
UMD-PP-96-93\\
SMU-PHYS-96-02\\
FERMILAB--Pub--96-096-A\\
hep-ph/9605350\\
May, 1996}
\end{flushright}

\vspace{2mm}

\begin{center}
 
{\Huge \bf New Supernova Constraints on Sterile Neutrino Production}\\

\vspace{1.3cm}
{\large Edward W.\ Kolb,$^{a}$  Rabindra N.\ Mohapatra,$^{b}$\\   
\vspace{6pt} 
and Vigdor L.\ Teplitz,$^{b,c}$}
\\ 
[5mm]
$^a${\em NASA/Fermilab Astrophysics Center,\\
Fermi National Accelerator Laboratory, Batavia, Illinois~~60510,  
and\\
Department of Astronomy and Astrophysics, Enrico Fermi Institute,\\
The University of Chicago, Chicago, Illinois~~60637}\\
[2mm]
$^b${\em Dept. of Physics, University of Maryland, College Park, 
Maryland 20742}\\ 
[2mm]
$^c${\em Permanent address:
 Department of Physics, Southern Methodist University, Dallas,
Texas~~75275}\\
[2mm]

\end{center}

\vspace{2mm}
\begin{abstract}
We consider the possibility that a light, sterile-neutrino species
$\nu_S$ can be produced by $\nu_e$ scattering during the cooling of a
proto-neutron star.  If we parameterize the sterile-neutrino
production cross section by a parameter $A$ as $\sigma(\nu_e X
\longrightarrow\nu_S X) = A \sigma(\nu_e X \longrightarrow\nu_e X)$,
where $X$ is an electron, neutron, or proton, we show that $A$ is
constrained by limits to the conversion of $\nu_e$ to $\nu_S$ in the
region between the sterile-neutrino trapping region and the
electron-neutrino trapping region.  This consideration excludes values
of $A$ in the range $10^{-4} \la A \la 10^{-1}$.
\end{abstract}
 
\newpage
\renewcommand{\thefootnote}{\arabic{footnote})}
\setcounter{footnote}{0}
\addtocounter{page}{-1}
\baselineskip=24pt

The possibility that the solar-neutrino problem \cite{bahcall} may be
solved via the oscillation of electron neutrinos to ``sterile''
neutrinos (so named because they have superweak interaction with the
$W$ and the $Z$ and thus avoid the LEP bound on the number of
neutrinos) has been widely discussed in literature.  The most
compelling case for sterile neutrinos \cite{calmoh} arises when one
tries to solve simultaneously the solar-neutrino problem and the
atmospheric-neutrino deficit, as well as accommodating either (or
both) the reported $\bar{\nu}_{\mu} \longrightarrow \bar{\nu}_e$
oscillations at LSND \cite{LSND} or the idea that neutrinos constitute
about 20\% of the total matter content of the universe \cite{HDM}.  In
the sterile-neutrino hypothesis the solar-neutrino deficit is resolved
via MSW oscillations between $\nu_e$ and $\nu_S$.

In constructing realistic gauge models \cite{calmoh,bm,smirnov} that
lead to the mixings between the electron neutrino and the sterile
neutrino one generally introduces various new interactions which can
lead to the desired mixing without conflicting with known low-energy
constraints as well as cosmological and astrophysical ones
\cite{juha}.  A well known astrophysical phenomenon that 
leads to strong constraints on the static properties of the neutrino is
the dynamics of supernovae inferred from the neutrino signal from
supernova 1987A (SN1987A) observed by the underground detectors of the
IMB and Kamiokande collaborations \cite{IMBKII}. Two classes of
restrictions on sterile neutrino properties may be obtained.  One is
on $\nu_e$ oscillating into $\nu_S$, thereby depleting the $\nu_e$
signal and contradicting observations. This possibility has been
analysed by Kainulainen et al.\ \cite{kimo}, who showed that the high
density in the supernova suppresses such oscillations for the range of
masses and mixing angles needed to solve the solar neutrino
problem. The other class of constraints may arise in models where
there exist {\em direct} interactions of electron neutrinos with
visible particles such as $e, p, n, \nu_{e,\mu,\tau}$ that can convert
a $\nu_e$ into a $\nu_S$. This can also potentially deplete the
$\nu_e$ luminosity. It is this class of effects that we discuss in
this letter.  We will also apply our techniques to restrict the mixing
between the photon and a hypothetical mirror (or para) photon.
   
The observed neutrino signal from SN1987A appears to in agreement with
expectations from the standard picture of type II supernovae
\cite{burrows}, with neutrino interactions of the standard model of
particle physics.  Any new interaction of the neutrino will therefore
be constrained by these observations. Some examples of constraints
already discussed in the literature are limits on the magnetic moment
\cite{mag}, the strength of right-handed interactions \cite{raffelt},
and the magnitude of the Dirac mass \cite{gandhi} of the neutrinos
\cite{revraf}. Similar considerations can be applied to new
sterile-neutrino interactions.  In this paper we study limits to the
production of sterile neutrinos in electron-neutrino collisions with
normal matter.

Before starting we emphasize that there are two different ways of
producing sterile neutrinos in the supernova: (i) $\nu_e$-$\nu_S$
mixing, which can convert electron neutrinos already in the supernova
to sterile neutrinos via oscillations; and (ii) direct production of
sterile neutrinos in the electron neutrino collisions with matter in
the supernova.  We will be concerned only with the second one since
the first effect has been shown by Kainulainen et al.\ \cite{kimo} to
be unimportant for our range of parameters due to MSW suppression.

A simplified model of neutrino interactions in the proto-neutron star
will be adequate for our purposes.  We assume that the core consists
of a sphere $10^6$cm in radius at a constant temperature of about 10
MeV, and density approximately equal to that of nuclear matter, $\rho
= 3\times 10^{14}$g cm$^{-3}$. The neutrino scattering cross section
is roughly $\sigma_{ee} \simeq G_F^2E_\nu^2$.\footnote{A matter of
notation: by $\sigma_{ij}$ we mean the cross section for $\nu_i + X
\longrightarrow \nu_j + X$, where $X$ is a normal matter particle.
For example, $\sigma_{ee}$ is the cross section for $\nu_e + X
\longrightarrow \nu_e + X$, while $\sigma_{eS}$ is the cross section
for $\nu_e X \longrightarrow \nu_S X$.}  Using $E_\nu=3T$, the
mean-free-path of the neutrino is about $\lambda_e \sim 10^2$cm.  This
means that the neutrino random walks out of the core, taking on
average $(R_C/\lambda_e)^2\sim 10^8$ steps.  So a typical neutrino
travels $10^8\lambda_e \sim 10^{10}$cm through the core, requiring
about a second.

Since electron neutrinos are trapped, effectively they are emitted
from a neutrinosphere, analogous to the familiar photosphere, where
the optical depth for a neutrino travelling out of the core is unity.
Observations by IMB and KII of neutrinos from SN1987A verify the
prediction of Colgate and White \cite{Stirling} that almost all the
binding energy of the neutron star is emitted in the form of
neutrinos.  Furthermore, a fair fraction must have been in the form of
$\nu_e\bar{\nu}_e$ pairs.  Thus, if there are additional weakly
interacting particles produced under the conditions of the
proto-neutron star, they cannot modify the fact that a significant
fraction of the binding energy must be radiated in the form of
$\nu_e\bar{\nu}_e$ pairs.  For instance, if the energy loss due to a
new hypothetical particle were too rapid, then the core would cool too
rapidly without emitting the observed neutrinos, leading to a conflict
with observation.  Or if a process somehow prevented
$\nu_e\bar{\nu}_e$ emission, then that process would be disallowed.

If sterile neutrinos are produced in the core, then in order that they
not carry away a disproportionate share of the binding energy, we must
either require that they are hard to produce, or else require that it
be difficult for them to escape. Since the sterile-neutrino
mean-free-path is $A^{-1}$ larger than that for the electron neutrino,
trapping will obtain for $A > 10^{-4}$.

Let's first examine the situation where the sterile neutrinos are
trapped.  If they are trapped and form their own neutrinosphere, the
ratio of the electron-neutrino luminosity to the sterile-neutrino
luminosity would be $r_\nu \equiv {\cal L}(\nu_e) / {\cal L}(\nu_S) =
R_e^2 T_e^4/R_S^2 T_S^4$ where $R_e$ ($R_S$) and $T_e$ ($T_S$) are the
radius and temperature of the electron-neutrino (sterile-neutrino)
neutrinosphere.  If the sterile neutrinosphere is deep in the core the
temperature will be higher, but let's assume for a moment that the
temperature is the same as the electron-neutrinosphere.  From the
universality of the weak interactions we know that $A$ must be much
less than unity in realistic models \cite{calmoh,bm,smirnov}, e.g.,
the effective Fermi constant for sterile neutrinos must be smaller
than $G_F$ from the fact that the sterile neutrino interactions
generally contribute additional modes to muon and tau lepton decays.
Since we expect $A<1$ from low-energy weak-interaction data, we would
have $R_S\la R_e$.  Therefore, so long as sterile neutrinos are
trapped, the emission {\em from the sterile neutrinosphere} will be
less than that from the electron neutrinosphere. That is, if $10^{-4}
\la A $, then neutrinos will be trapped, and radiation from the
sterile neutrinosphere will result in a sterile-neutrino luminosity
less than the electron-neutrino luminosity.  The point of this paper,
however, is to observe that the sterile-neutrinosphere would {\em not}
be the dominant source of sterile neutrinos.

Now consider the possibility that sterile-neutrino interactions are so
feeble that they are not trapped.  Then one must limit the production
of the sterile neutrinos.  This results in $A\la 10^{-10}$.  An easy
way to see the origin of this bound is to note that the sterile
neutrino luminosity ${\cal L}(\nu_S) $ in the non-trapped case is
directly given by the total number of $\nu_S$'s produced in the
supernova core times the average neutrino energy.  This is given by
\begin{equation}
{\cal L}(\nu_S) \simeq n_e n_{\nu_e}A\sigma_{ee} V \langle E \rangle,
\end{equation}
where $n_i$ represent the number density of relevant particles and $V$
is the volume of the supernova core.  Using $n_i\simeq
10^{38}$cm$^{-3}$ and $E=3T$ with $T\simeq 50$ MeV, and demanding that
${\cal L}(\nu_S) \leq 10^{53}$ erg s$^{-1}$, we obtain $A\la
10^{-10}$.

So to review the {\it standard analysis}, values of $A$ in the range
 $10^{-4}\la A $ result in a sufficiently small luminosity from
the sterile neutrinosphere because of the small trapping radius.
Values of $A$ in the range $10^{-4}$ to $10^{-10}$ are not allowed
because in this regime the sterile neutrinos free stream and have a
sufficiently large production cross section to be dangerous.  Thus,
the usual supernova analysis allowed regions for $A$ are 
$ 10^{-4}\la A$, and $A\la 10^{-10}$.

We, however, find that if the new interactions that convert $\nu_e$'s to
$\nu_S$'s are strong enough to satisfy the trapping criteria, then a
new consideration appears to lead to more stringent limits on the
strength of these interactions than one would obtain using the
familiar arguments discussed above \cite{revraf}.  This new
consideration will exclude $A\ga 10^{-4}$, so the final result will be
that the only allowed range of $A$ is $A \la 10^{-10}$.

Consider the trapped sterile-neutrino scenario in the ``flat-star''
approximation \cite{brod}, i.e., as a one-dimensional problem.  We
know that the sterile neutrinosphere is well within the electron
neutrinosphere.  Consider the fate of an electron neutrino between the
two neutrinospheres.  Let $n$ be the density of scatterers ($e, n, p,
\nu_{e,\mu,\tau}$) and $\sigma_{ij}$ the cross section for $\nu_i$
scattering into $\nu_j$ as before.  We assume that, for the sterile
$\nu_S$, $\sigma_{SS}$ is negligible.    The $\nu_e$ and $\nu_S$
mean-free-paths are then given by
\begin{equation}
\lambda_{ee}=1/(n\sigma_{ee}) ; \qquad  \lambda_{eS}=1/(n\sigma_{eS})
=\lambda_{ee}/A .
\end{equation}
Near $R$, the radius of the SN core, let $R_e=R-\lambda_{ee}$ and
$R_S=R-\lambda_{Se}.$ So long as $\lambda_{ee}$ and $\lambda_{Se}$ are
much less than $R$, the relation
\begin{equation}
dn_i/dt=-n_ic\sigma_{ij}+n_jc\sigma_{ji}\sim 0
\end{equation}
gives that $n_S \sim n_e$ at $R_S.$ Thus $1/(2e)$ of neutrinos passing
$R_S$ will exit promptly.  For $R>R_S$, we will have that $\nu_e$
scattering into $\nu_S$ with $\nu_S$ exiting without further
scattering depletes the number of $\nu_e$'s. A $\nu_e$ traveling a
distance $\lambda_S$ without changing to a $\nu_S$ will have suffered
$1/A^2$ scatterings. The square is because of the random-walk nature
of the path.  The chances of a $\nu_e$ surviving so many scatterings
without changing to a $\nu_S$ are
\begin{equation}
P(R_S)=[1-A/(1+A)]^{1/A^2} \simeq \exp[-1/A(1+A)] .
\end{equation}
As $A$ goes to zero, this result fails when $\lambda_{eS}$ approaches
$R$ since the number of scatterings stops decreasing continuously
below one; as $A$ approaches one it remains qualitatively correct.
Thus, except for $A$ very close to one, essentially no $\nu_e$ survive
the trip from $R_S$ to $R.$ All exiting $\nu_e$ must be the result of
either ``local production'' ($\nu_e$ absorption followed by $\nu_S$
emission can be considered incorporated in $\sigma_{eS}$), or
``regeneration.'' We now compute the fraction from regeneration.

For a $\nu_S$ approaching $R$. the chance of it scattering into a
$\nu_e$ in a length $dx$ at a distance $x$ before $R$ is
$dx/\lambda_{eS}.$ The chance of the $\nu_e$ produced surviving the
distance from $x$ to $R$ is
\begin{equation}
P(x)=[1-A/(1+A)]^{x^2/\lambda_{ee}^2} \sim
\exp[ -x^2/(1+A)\lambda_{ee}\lambda_{eS}] .
\end{equation}
The fraction, $f$, of exiting $\nu_e$'s is then the product of these two 
probabilities summed over distances $x$:
\begin{equation}
\label{eq:six}
f=\int P(x)dx/\lambda_{es}=\sqrt{\pi A(1+A)}\, /2 \sim \sqrt{A} .
\end{equation}
Thus, if the coupling constant to the sterile neutrino is $1/3$
that to the electron neutrino ($A=0.1$), only one-third of the
exiting neutrinos will be electron neutrinos.  As a result, the range
in the parameter $A$ for which a ``sterile'' neutrino can be confined
in a supernova and permit a reasonable number of $\nu_e$ to exit is
limited to $A$ close to one, roughly $A>0.1$.

There are several possible corrections to this result.  They include:
(i) production of $\nu_e$ by $\nu_{\mu}$ or $\nu_{\tau}$ pairs; and
(ii) MSW \cite{MSW} oscillations which may regenerate the electron
neutrinos from the sterile neutrinos as they pass through the dense
remainder of the neutrino sphere.  As for possibility (i), one may
show by arguments similar to those above that, because the
$\nu_{\mu}$ and $\nu_{\tau}$ mean-free-paths are larger than that of
$\nu_e$, they tend to decrease the $\nu_e$ flux as the $\nu_S$ does,
but because neutrino densities are small compared to matter density
near $R_S$, the effect is small.  Let us briefly comment on the second
aspect. As the converted sterile neutrinos pass through the neutrino
sphere (or what is left of it after they are produced), they
experience a matter potential due to MSW effect which differs from
those of the $\nu_e$ and $\nu_{\mu}$ as follows:
\begin{eqnarray}
V(\nu_S) & = & 0 \nonumber \\
V(\nu_e) & = & V_0(3Y_e-1+4 Y_{\nu_e}) \nonumber \\
V(\nu_{\mu}) & = & V_0(Y_e-1+2Y_{\nu_e}),
\end{eqnarray}
where $V_0=18~{\rm eV}\left(\rho/5\times 10^{14}{\rm
g~cm}^{-3}\right)$, the factors of $Y$ represent the fraction of the
corresponding species relative to the total number of nucleons
\cite{kimo}, and $\rho$ is the density.  The MSW resonance therefore
may occur if the expression within the parentheses above vanishes.
If it does vanish, one may have conversion of $\nu_S$ to $\nu_e$ if the
latter are lighter. It is hard to estimate the precise number of the
$\nu_S$'s that would be converted. And in any case, this will require
accidental fine tuning of the particle densities at the right distance
from the surface of the supernova.

Let us now discuss the implications of our result:

\noindent(i) In order that $\nu_e\bar{\nu}_e$ pairs be emitted from
the supernova, if sterile neutrinos interact strongly enough to be
trapped, the sterile neutrinosphere must be close to the electron
neutrinosphere.  This requires $A\ga 10^{-1}$, which would be in
conflict with what we know about the weak interactions.  In
particular, we know that $\nu_e$--$e$ neutral current scattering
agrees with the standard-model result to a few percent.  Setting a
precise limit to $A$, however, would require a specific
sterile-neutrino model.  Thus, the true bound on $A$ is the upper
bound derived from luminosity discussion in SN1987A, i.e., $A\la
10^{-10}$.

\noindent(ii) Our result also has several implications for models
incorporating sterile neutrinos that must be ultra light. Any model
that has effective Four-Fermi interactions of $\nu_S$ with $e,\nu_e,
\nu_{\mu}$, etc., with strength above $G_F\times 10^{-5}$, will be
ruled out. It is interesting that the mirror model \cite{bm} for
sterile neutrino is among the models that are consistent with the
above constraints, since all interactions between the visible sector
particles and the $\nu_S$'s in this model are Planck-scale
suppressed. The two models \cite{calmoh} that use two-loop graphs to
suppress the $m_{\nu_S}$'s generically involve larger couplings but
use $A\simeq 10^{-10}$ so that they are also barely consistent with
these constraints.  On the other hand several models constructed to
explain the 17 KeV neutrino had larger $\nu_s$--$e$ cross-sections
with $A$ in the range of $10^{-4}$ or so and are inconsistent with our
improved supernova limit.

\noindent(iii) The above bound is independent of the mass of the
sterile neutrino as long it is light enough to be produced in
supernova temperatures (say, $m \la10$ MeV or so).

\noindent(iv) Another application of our bound is to the Dirac
magnetic moment of the tau neutrino with mass in the MeV range, which
has sometimes been considered in literature to be large so that it
could be the dark matter of the universe \cite{giudice} or have an
effect on big-bang nucleosynthesis \cite{grasso}. In this case, if the
magnetic moment is larger than $10^{-8}\mu_B$, Giudice showed that the
mean free path for $\nu_{\tau}$ will be {\it less} than that of
$\nu_e$, and the $\nu_e$ neutrinosphere will be within the $\nu_\tau$
neutrinosphere.  Hence, by Eq.\ (\ref{eq:six}), $\nu_{\tau}$ and
$\bar{\nu}_\tau$ will tend to be converted to $\nu_e$ and
$\bar{\nu}_e$ (as well as $\nu_{\mu}$ and $\bar{\nu}_\mu$) while
traversing the region between the neutrinospheres. This would enhance
the low energy $\nu_e$ and $\bar{\nu}_e$ signal in the underground
detectors.

\noindent(v) A final implication of our discussion is that 
one could apply the techniques
of our paper to constrain the mixing of of a new photon \cite{holdom}
with the known photon using the supernova luminosity information. The
discussion is very similar to the case of neutrinos.  Let us consider
only the effect of photon scattering off electrons.  If we denote the
shadow photon as $\gamma'$ and the $\gamma-\gamma'$ mixing to be
$\epsilon$, then the rate of supernova cooling via $\gamma'$ emission
per unit volume is given by:
\begin{equation}
Q_{\gamma'}= n_e \int {{dn_{\gamma}}\over{d\omega}}\, 
\sigma _{\gamma\gamma' }\,
 \omega  \ d\omega  \ .
\end{equation}
The cross section for $\gamma + e \longrightarrow \gamma' + e$ is just
$\epsilon^2$ times the Compton cross section:
$\sigma_{\gamma\gamma'}=\epsilon^2\pi \alpha^2/2\omega^2$. Integrating
over the photon spectrum, one finds for the total luminosity via the
$\gamma'$ channel to be
\begin{equation}
Q_{\gamma'}={{ \alpha^2\epsilon^2 }\over{6\pi }} n_e \int 
{{\omega d\omega}\over{e^{\omega/T}-1}} 
= \frac{\alpha^2\epsilon^2 \pi}{36} n_e T^2.
\end{equation}
The supernova luminosity is obtained from this by multiplying the
volume of the supernova. Assuming $T\simeq 50 $ MeV as the temperature
of the supernova core, and the number density of electrons in the
supernova to be $1.5\times 10^{38}$cm$^{-3}$, we get
\begin{equation}
Q_{\gamma'}V\simeq 10^{72}\epsilon^2~~\ergs .
\end{equation}
Demanding that this be less than $10^{53}\ergs$, we obtain the bound
$\epsilon\leq 10^{-9.5}$.  In principle in the supernova as well as
the solar case, there would have been a region close to $\epsilon\geq
10^{-8}$ where trapping arguments would have said that shadow photon
is allowed. However our discussion here can be applied to the
supernova after breakaway when the first electromagnetic signals were
observed\footnote{Similarly, Primakoff process conversion of photons
to axions should lead to constraints, although necessarily new ones,
on the axions.}. It is simpler, however, just to consider the sun. If
$\epsilon = 0.5$ or less, three quarters or more of the solar
luminosity would be in para-photons ($\gamma'$). This essentially
rules out the region $\epsilon \geq 10^{-9.5}$. We note that
$\epsilon$ would exhibit itself as a minicharge of the same magnitude
on mirror electrons and a recent experiment at SLAC
\cite{prinz} has ruled out minicharged particles above charge $7\times
10^{-5}e$.

In conclusion, from considerations of the electron neutrino luminosity
from SN 1987A we have pointed out a new effect that considerably
limits the allowed couplings of the sterile neutrinos to ordinary
matter.

We thank G. Snow for several conversations and for bringing the last
reference to our attention. One of the authors (R.N.M.) would like to
thank K. Kainulainen for discussions and CERN theory group for
hospitality when the paper was completed.  The work of E.W.K.\ was
supported by the DOE and NASA under grant NAG5-2788. The work of
R.N.M.\ is supported by the National Science Foundation and a
Distinguished Faculty Research Award from the University of Maryland.
It is a pleasure to acknowledge the use of computing facilities
provided by the Lightner-Sams Foundation.

\end{document}